\documentclass[useAMS,usenatbib]{mn2e}

\usepackage{graphicx,epsfig}
\usepackage{ulem}

\def\be{\begin{equation}}
\def\ee{\end{equation}}
\def\ba{\begin{eqnarray}}
\def\ea{\end{eqnarray}}
\def\de{\partial}
\def\12{{1\over 2}}

\def\msun{M_\odot}

\def\etal{{\it et~al.~}}
\def\ltsima{$\; \buildrel < \over \sim \;$}
\def\simlt{\lower.5ex\hbox{\ltsima}}
\def\gtsima{$\; \buildrel > \over \sim \;$}
\def\simgt{\lower.5ex\hbox{\gtsima}}

\title[Probing primordial $^3$HeII by supercritical BHs]
{Probing primordial $^3$He from hyperfine line afterglows around supercritical black holes}
\author[E. O. Vasiliev \etal]
       {Evgenii O. Vasiliev$^{1,2}$\thanks{E-mail:eugstar@mail.ru},
        Shiv K. Sethi$^{3}$,
        Yuri A. Shchekinov$^{2,3}$\\
$^1$Southern Federal University, Stachki Ave. 194, Rostov-on-Don, 344090 Russia\\
$^2$Lebedev Physical Institute of Russian Academy of Sciences, 53 Leninskiy Ave., 119991, Moscow\\
$^3$Raman Research Institute, Sadashivanagar, Bengaluru 560080, Karnataka, India
}
\begin{document}
\date{Accepted 2018 December 32.
      Received 2018 December 31;
      in original form 2018 December 31}
\pagerange{\pageref{firstpage}--\pageref{lastpage}}
\pubyear{2119}
\maketitle

\label{firstpage}

\begin{abstract}
We consider the possibility of the detection of $^3$HeII hyperfine line (rest frequency, $8.67 \, \rm GHz$) emission from ionized zones around  accreting black holes (BHs) formed at high redshifts, $z=15\hbox{--}30$. We show that the brightness temperature in 8.67GHz line increases and reaches a peak value after the accretion onto the BH exhausts and  HeIII recombines into HeII. This period of brightening last up to  40~million years. We find that during this period the maximum brightness temperature reaches  $\simeq 0.2\hbox{--}0.5 \mu$K, depending on the epoch when such a black hole starts growing. The maximum angular size of the region emitting in the hyperfine line  is around $0.5'$. The flux from such a region ($\simeq 0.3 \, \rm  nJy$) is too small to be detected by SKA1-MID. The RMS of the  collective flux from many  emitting regions from a volume bounded by the synthesized beam and the  band-width of SKA1-MID might reach 100~nJy, which is potentially detectable by SKA1-MID. 
\end{abstract}


\section{Introduction} 

The primordial  Nucleosynthesis (BBN)  is one of the pillars of the hot big bang model. While the initial success of this proposal was based on the formation of $^4$He, the detection of these elements in their primordial settings continues to be a challenge, which is  particularly severe for  the small amounts of  trace elements---$^2$D, $^3$He and $^7$Li---produced during the  BBN. 

The detection  of line absorption in the  spectra of distant quasars has long been recognized to be a powerful tool for probing the  chemical evolution of the universe across the Hubble time \citep[see review in][]{procha17}. In this series the direct observation of deuterium Ly-series line absorption from redshifts $z=2\hbox{--}3$ plays an exceptionally important role as a unique probe of the Big-Bang nucleosynthesis \citep[BBN -- see for recent  discussion in][]{balash16,cook18,zavar18}. Such observations though are biased  towards damped Ly-$\alpha$ absorption systems (DLA) with matter partly processed by stellar nucleosynthesis, and hence the necessity to reach deeply pristine conditions urge search of rare DLA systems in the lowest end of metallicities like reported by \cite{cook18} with [O/H]=-2.769. Among other complications, the   rareness of such systems can make conclusions elusive  \citep[e.g.][]{cyburt16,draine06}. The  primordial abundance of $^7$Li and $^4$He has also not been determined unambiguously \citep{2011ARNPS..61...47F,izot14,aver15,cooke15}. 

The focus of the current paper is $^3$He,  with BBN abundance of nearly  $10^{-5}$. One possible approach to measure the  primordial abundance of $^3$He is  through the simultaneous  measurement of deuterium abundance and the isotope ratio $^3$He/$^4$He from metal-poor HII regions. This would require  using future   optical and near-infrared observational facilities of  30m class telescopes  \citep[][]{cooke15}. An alternative option to measure the BBN abundance of $^3$He is to  detect the absorption owing  to  the hyperfine  $^3$HeII line \citep{townes57,sunyaev66,goldwire67,rood79} in the quasar spectra at  $z\simgt 3$ \citep[][]{mcquinn09}. { The $^3$HeII  hyperfine line has been detected from local HII regions and planetary nebulae using both single-dish radio telescopes (e.g. GBT) and radio interferometers (e.g. VLA) \citep[e.g.][]{2018AJ....156..280B,2010IAUS..268...81B,2006ApJ...640..360B}}. 

More recently \citet{takeuchi14} studied the possibility of detecting the hyperfine line in emission from the intergalactic low-overdensity  gas exposed to the cumulative ionizing background at $z\simlt 10$ \cite[see also][]{bell00,bagla09}. In this paper, we study the emission of this line from ionized regions surrounding growing black holes in the redshift range $12 <z<25$ and investigate its detectebility using the upcoming radio interferometer SKA1-MID. 

Throughout this paper, we assume the spatially-flat $\Lambda$CDM model with the following parameters: $\Omega_m = 0.254$, $\Omega_B = 0.049$, $h = 0.67$ and $n_s = 0.96$, with the overall normalization corresponding to  $\sigma_8 = 0.83$ \citep{Planck2015}. The helium abundance is  $Y_{\rm ^4He}=0.24$.

\section{THE GROWTH OF BH} \label{sec:bhgrowth}

A black hole (BH) is assumed to grow with the Bondi-Hoyle rate \citep{bondi,bondi52,shima85}:
\ba
 \dot M_{Bondi} = {4\pi G^2 M_{\rm BH}^2 \mu m_{\rm H} n \over c_s^3 } = 4\times 10^{-5} 
                                                      \left( {M_{\rm BH} \over 10^3 \msun}\right)^2 \times  \nonumber \\ 
   \times \left( {n_0 \over 10^4~ {\rm cm^{-3}}} \right) \left( {T_0 \over 8000~{\rm K} } \right)^{-1.5} ~ \msun~{\rm yr^{-1}} \ \ \ \
\label{dotm}
\ea
where $n_0$ is the density of the gas at the center of the gaseous disks in host halos. 
\footnote{{ The  ``host halo'' refers to a  virialized object  with the overdensity $\Delta_s=18\pi^2$ with respect to the background density at that redshift, with baryon-to-dark matter ratio $\Omega_B/\Omega_{m}=0.19$ \citep{barkana01}. } }
The bolometric luminosity equals 
\be
 L_{BH} = \eta \dot M_{Bondi} c^2
\label{lumbh}
\ee 
where $\eta = 0.3$ is assumed as usual; it  corresponds to the case of radiation from a thin disk around the BH \citep{ss73}. Note that the radiative efficiency $\eta$ may vary during the accretion history in different models, which mimic the effect of photon trapping. Depending on ambient gas parameters and the BH mass its luminosity may vary from slightly subcritical to highly supercritical 
\be \label{edreg}
   {\L_{BH}\over L_{\rm Edd}}\simeq 18\eta\left( {M_{\rm BH} \over 10^3 \msun}\right)\left( {n_0 \over 10^4~ {\rm cm^{-3}}} \right) \left( {T_0 \over 8000~{\rm K} } \right)^{-1.5},
\ee
and can reach up to hundreds of $L_{\rm Edd}$ \citep[see e.g.][]{inayoshi16,sakurai16}.

\section{The $^3$HeII hyperfine transition} \label{sec:model}

The $^3$HeII hyperfine transition ($\nu = 8.666, \rm GHz$) is excited by collisions with atoms and electrons and photon scattering \citep{field,wout} 

\be \label{tspin}
T_s^{\rm ^3HeII} = {T_{CMB} + y_c T + y_a T\over 1 + y_c + y_a}, 
\ee 
where $y_c=(T_\ast/T)(C_{10}/A_{10})$ is the collisional coupling efficiency of the two states of hyperfine splitting, $C_{10}=n_ev_{e}14.3~{\rm eV}~ a_0^2/k_{\rm B}T$ is the rate of spin exchange between $^3$HeII and electrons \citep{mcquinn09}, $A_{10}$, the rate of spontaneous decay of the hyperfine state, $k_{\rm B}T_\ast=h\nu_{10}$ is the energy of the hyperfine splitting; $y_a$ denotes the efficiency of Wouthuysen-Field coupling by photons of wavelength, $\lambda = 304 \, \rm \AA$, corresponding to the Lyman-$\alpha$ transition energy for singly ionized helium \citep[Eq.~17 in][]{takeuchi14}, $a_0$ is the Bohr radius. In this paper, the number density of these photons at the point of scattering is computed from the spectrum of BH emission \citep[for more details see in][]{vss18}. 

The differential brightness temperature of $^3$HeII line can be estimated as follows:
\ba
   \Delta T^b_{^3{\rm HeII}} = 1.7~{\rm \mu K} ~ (1 + \delta) ~  X_{\rm HeII}  \left( {Y_{\rm ^3He}\over 10^{-5}}\right)  \times \nonumber \\
                             \times  ~  {T_s^{\rm ^3HeII} - T_{CMB} \over T_s^{\rm ^3HeII}} 
                      \left({\Omega_b h \over 0.03}\right)    \left({0.3 \over \Omega_m}\right)^{0.5}   \times \nonumber \\
               \times    \left({1+z\over 10}\right)^{0.5}                 
                     \left[ { H(z)/(1+z) \over dv_{||}/dr_{||} }\right]
\label{tbhe}
\ea
where $Y_{\rm ^3He} = 10^{-5}$ is the primordial abundance of the helium-3 isotope, $X_{\rm HeII}$ is the fraction of singly ionized helium-4 isotope; as the chemical properties of both the isotopes are nearly the same\footnote{They differ only owing to the slightly different reduced masses of the atoms formed with these nuclei.}, $X_{\rm HeII}$ is also the fraction of singly ionized helium-3.

\section{The ionized IGM around BH} \label{sec:resu}

The accretion onto a BH is expected  to be  a source of UV/X-ray photons. The spectrum of the ionizing radiation emitted during the accretion is assumed  a power-law $L_\nu \sim \nu^{\alpha}$, where $\alpha = -1.5$, which is taken  as a fiducial value here. This ionizing radiation is attenuated inside the halo by average total column density of HI, $N_{\rm HI}^{h}$, in  the host galaxy. After escaping the halo the radiation can ionize the intergalactic medium. We neglect the complications of the transition layer between minihalo and surrounding gas and match the minihalo 
directly to the expanding IGM. We assume this gas to have homogeneous distribution of density and temperature decreasing due to cosmological expansion:  $\propto (1+z)^{-3}$ and $\propto (1+z)^{-2}$,  until the ionizing radiation from a BH  changes its thermodynamics. We consider the evolution of gas enclosed in the concentric static spheres with a BH in the center. In each sphere we solve thermal and ionization evolution of hydrogen, neutral helium and singly ionized helium. We consider all major processes for primordial plasma  \citep{cen92}. We include the influence of secondary electrons as described in \citep{steenberg85,ricotti02}. In the  equation of the thermal evolution of IGM  we include  the cooling term due to the Hubble expansion, in order to correctly describe the evolution on time scales greater than the local age of the universe. The initial gas temperature and HII fraction for a given redshift are  obtained by using the RECFAST code \citep{recfast}, while helium in the initial state is assumed to be neutral. The detailed description of our model can be found in \citep{vss18}.

Based on that we consider BH hosted in halos at redshifts $z\simgt 10$ with the BH seeds being the final product of the evolution of Population III stars with $M \sim 30\hbox{--}260\msun$  \citep[see e.g.,][]{woosley02}, $M_{BH}=300~\msun$ is adopted as a fiducial value for a BH seed. 

\begin{figure}
\center
\includegraphics[width=78mm]{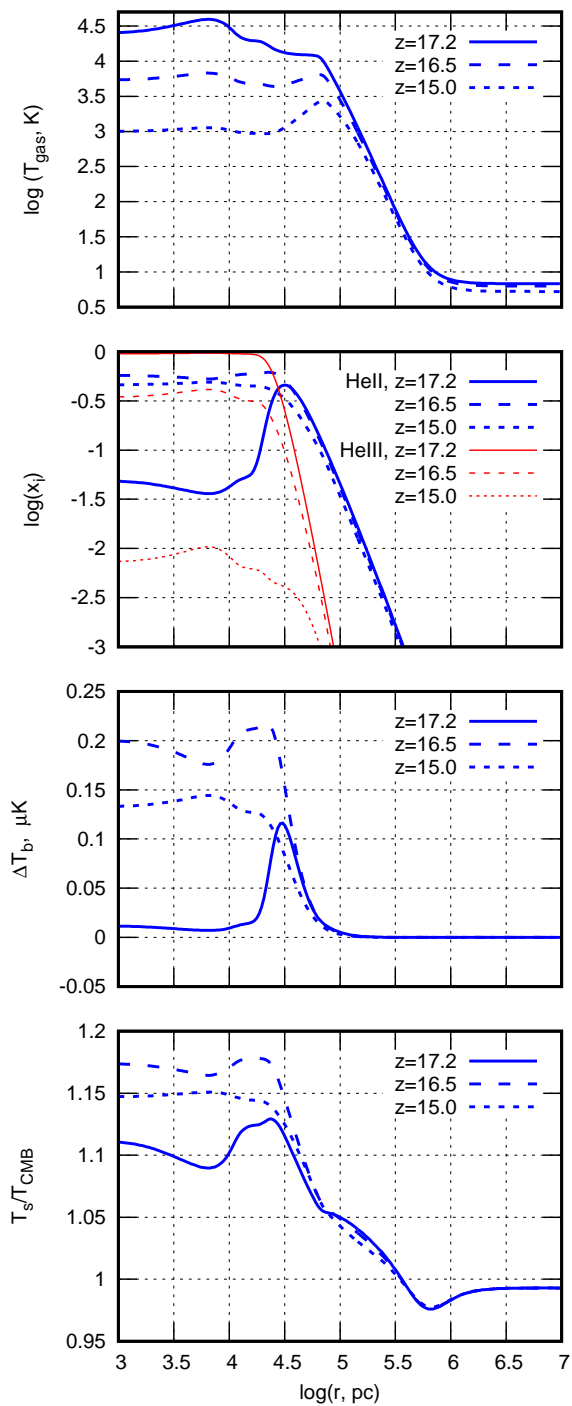}
\caption{
The radial distribution (radius is in physical, not comoving units) of the kinetic temperature (upper), HeII and HeIII fractions (middle), $^3$HeII brightness temperature (lower middle), $T_s^{\rm ^3HeII}/T_{\rm CMB}$ (lower) around a BH with initial mass $M_{BH}(z_0)=300\msun$ starting its evolution at $z_0 = 20$ and increasing the mass by a factor of  1000 until  redshifts: $z = 17.2$ (solid line) when accretion completes, with the corresponding mean luminosity $\simeq 3L_{\rm Edd}$; and radial profiles for intermediate redshifts $16.5,\,15.0$ (dashed and dotted lines, respectively). The BH grows with a single accretion episode. The disk around the BH has number density $n_0 =10^4$~cm$^{-3}$ and temperature $T_0=8\times 10^3$~K (see eq.~\ref{dotm}). 
}
\label{fig-evol}
\end{figure}

In order to illustrate the results we show several models with a black hole of  initial mass $M_{BH}(z_0)=300\msun$ which starts growing at $z_0 = 20$ and  increases  its mass by a factor of 1000 until   $z =  17.2$. Figure~\ref{fig-evol} presents the radial profiles of the  kinetic temperature, HeII and HeIII fractions, $^3$HeII brightness temperature around such a growing BH for several redshifts: $z = 17.2,\ 16.5,\ 15.0$ (solid, dashed and dotted lines, respectively). The BH grows through a single accretion episode completed at $z_e \simeq 17.2$. The disk around BH has number density $n_0 =10^4$~cm$^{-3}$ and temperature $T_0=8\times 10^3$~K (see Eq.~\ref{dotm}). At the end of the accretion episode, $z\simeq 17.2$, the IGM around BH is heated up to $T\sim (1\hbox{--}3)\times 10^4$~K within a sphere of radius $r\sim 100$~kpc (the upper panel of Figure~\ref{fig-evol}). The IGM is exposed to  the far-UV/X-ray ionizing radiation emitted by accreted gas and almost all the  helium is in the  doubly ionized state, HeIII,  up to distances $r\sim 20$~kpc (red solid line in middle panel). At larger distances, $r\sim 20\hbox{--}50$~kpc, there is a spherical layer, in which the singly ionized helium, HeII,  reaches a relatively high fraction ($x_{\rm HeII}\simeq 0.36$, see blue solid line in middle panel). This layer  emits in the $^3$HeII hyperfine structure line: the brightness temperature reaches 0.1~$\mu$K (see solid line in lower panel). The brightness temperature of the $^3$HeII line in the inner part of the ionized sphere is less than 0.01~$\mu$K. The lowest panel displays the ratio of $T_s^{\rm ^3HeII}/T_{\rm CMB}$ which shows that  the maximum value of this ratio is around 1.15. It should be  noted that  when the  gas cools the hyperfine spin temperature increases because the collisional coupling, $y_cT\propto T^{-1/2}$, gets stronger. Therefore, as soon as the cooling time  of the ionized gas $t_c\sim 10^{18}(1+z)^{-3}$~s  is shorter than the Hubble time $t_H(z) \sim 5\times 10^{17}(1+z)^{-3/2}$~s, which occurs immediately after the completion of  the accretion phase,  we notice an increase in the hyperfine spin temperature $T_s^{\rm ^3HeII}$. The collisional coupling  scales with $z_0$,  the redshift corresponding to the exhaustion of accretion on to growing massive BH, as $y_cT\propto (1+z)^3$. As a consequence, the resulting brightness temperature is  higher for larger  $z_0$. 

It follows from Eq.~(\ref{tspin}) that when  either $y_a \gg 1$ or $y_c \gg 1$, which denote  Lyman-$\alpha$ or collisional couplings, the spin temperature would approach the matter temperature $T$ which is generally much larger than $T_{\rm CMB}$ in the regions that emit in the hyperfine line. The collisional coupling depends on the number density of electrons in the IGM which is not high enough to drive $y_c$ to a value above unity. The Lyman-$\alpha$ depends on the number density of $\lambda = 304 \, \rm \AA$ photons in the IGM. In our study, we consider the BH to be the source of these photons. Most of these photons are strongly absorbed in the halo of HI column density $N_{\rm HI} = 10^{20} \, \rm cm^{-2}$ surrounding the BH, which diminishes their number in the IGM. As  $N_{\rm HI}$ is a parameter in our model, we studied many cases where it is smaller. For  $N_{\rm HI}= 10^{18} \, \rm cm^{-2}$, a much larger number of these photons escape into the IGM, greatly increasing $y_a$. However, this scenario also results in the escape of photons which can ionize the singly ionized helium, which diminished the hyperfine line signal. We find the net impact of changing $N_{\rm HI}$ to be negligible on the observable signal. In sum:   neither the collisional nor the Lyman-$\alpha$ coupling are  very efficient in coupling the spin temperature to matter temperature  which results in a decrement in the observed brightness temperature (Eq.~(\ref{tbhe})).

\begin{figure*}
\center
\includegraphics[width=175mm]{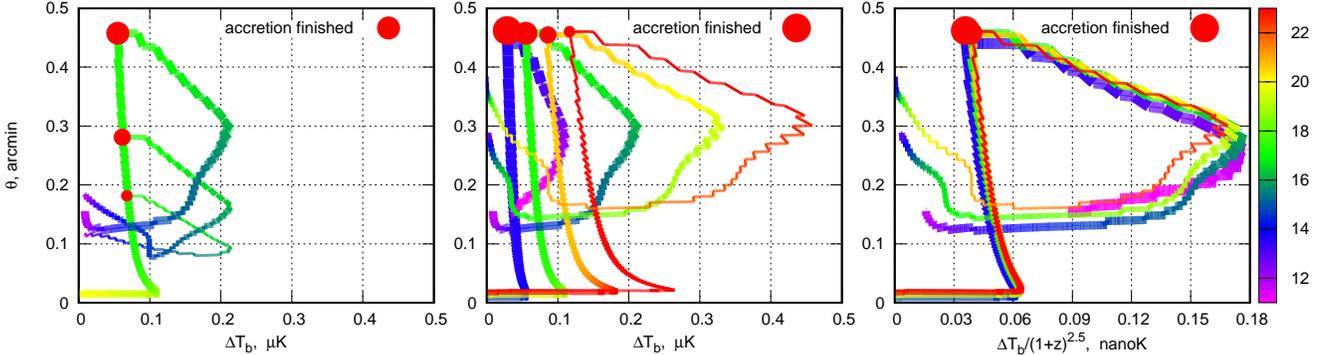}
\caption{
The evolutionary tracks of the maximum of $^3$HeII brightness temperature and the angular size of emitting sphere or the '${\rm max}[\Delta T_b] - \theta$' tracks. Redshift is indicated by color along the tracks. Red points depict the moment, at which the accretion onto a BH has been completed. {\it Left panel.} The '${\rm max}[\Delta T_b] - \theta$' tracks for the IGM surrounding a BH with the initial mass $M_{BH}(z_0)=300\msun$ starting evolution at $z_0 = 20$ and increasing its mass by a factor of  100 (the thinnest line and the smallest red point), 300 (thicker line and larger red point), 1000 (the thickest line and the largest red point) times. In all the cases the accretion episode was completed at $z_e \simeq 17.2$. {\it Middle panel.} The '${\rm max}[\Delta T_b] - \theta$' tracks for the IGM surrounding a BH with initial mass $M_{BH}(z_0)=300\msun$ starting its evolution at $z_0 = 15, 20, 25, 30$ (the thickness of the line and the size of the symbol (red point) is decreased for higher initial redshift), and increasing its mass by 1000 times; in all cases the average luminosity $L_{BH}\sim 3 L_{\rm Edd}$. The accretion episode has been completed to $z_e \simeq 13.5$, 17.2, 20.6 and 23.6, respectively. {\it Right panel.} Same as in the {\it middle panel}, but with the brightness temperature divided by $(1+z)^{2.5}$.
}
\label{fig-tb-size}
\end{figure*}

After the accretion onto BH has been switched off, the doubly ionized helium begins to recombine  (see dashed lines in middle panel), and the brightness temperature of $^3$HeII line grows: it reaches around 0.2~$\mu$K and has an  almost flat distribution within $r\simlt 30$~kpc. At lower redshifts,  the gas temperature  decreases and  the brightness temperature of $^3$HeII also drops, but it remains higher than that during the  accretion phase (dotted line in lower panel). The duration of this luminous 'afterglow' phase is $\Delta t\sim 40$~Myr ($z\simeq 17.2\hbox{--}15$). It  is comparable to  the accretion period ($z\simeq 20\hbox{--}17.2$). This  40~Myr period is indicative of  the  recombination time of HeIII. This period ends when the  recombination of HeII  becomes important, resulting in a drop in the brightness temperature of the hyperfine line. 

It is worth noting that the physical size of the brightest  part of the luminous sphere in the IGM slowly diminishes with the velocity
\be 
 v_{i}\sim {T_i\over t_r}\left|{\de T\over \de r}\right|_i^{-1}, 
\label{velo}
\ee
where subscript `{\it i}' stands for `ionized', $T_i$ is characteristic temperature in the ionized region, the derivative is taken at the edge of the ionization zone. In Fig. \ref{fig-evol}  $r_i\simeq 10^5$ pc and  $t_r$ is the recombination time of HeIII at $r_i$. For temperature at  the ``edge'' approximated as $T\simeq T_0(r_i/r)^\beta$ one has $v_i\simeq r_i/(\beta t_r)$. At the end of HeIII recombination when the brightness temperature peaks, the initial size of the ionization region $R_i\simeq [3L_{p,24.6}/4\pi\alpha_{\rm HeII}(T)n_e]^{1/3}$ decreases  by a  factor $c_i\sim 0.5\hbox{--}0.6$, where $L_{24.6,p}$ is the photon luminosity with photons energy $h\nu\geq 24.6$ eV, $\alpha_{\rm HeII, HeIII}(T)$ is the recombination coefficient of HeII (HeIII) ions, $n_e$ the number density of electrons, and the  subscript `0' corresponds to the end of  the accretion period. It follows then that independent of the initial size of the ionized bubble, after completing the accretion episode,  the region decreases by the same factor  at the end of HeIII recombination phase(as readily seen in Fig. \ref{fig-tb-size}). An important point is that if the  BH's mass grows by a factor the region  is proportionally larger,  whereas the brightness temperature reaches comparable  values.

Then, it appears to be useful to understand how $^3$HeII brightness temperature relates to size of emitting sphere. Strictly speaking, we should compare the maximum of $\Delta T_b$ and the size of the sphere at which the $\Delta T_b$ value reaches maximum. However, because of the non-monotonic $\Delta T_b$ radial profiles (see e.g. dotted line in the lower panel of Figure~\ref{fig-evol}) we may underestimate the size of the sphere: for the profile at $z=15$ the maximum is  8~kpc, whereas the profile is almost flat up to 20~kpc. So we use $0.9{\rm max}[T_b(r)]$ in the figure  and find the outer radius at which the value of $0.9{\rm max}[T_b(r)]$ is reached.

Figure~\ref{fig-tb-size} presents the evolutionary tracks for $^3$HeII brightness temperature $\Delta T_b$ and angular size of the emitting sphere,  $\theta$. The small saw-tooth fluctuations seen in the figure  are owing to the spatial resolution which is determined by the number of concentric shells in our calculation. The disk around BH has number density $n_0 =10^4$~cm$^{-3}$ and temperature $T_0=8\times 10^3$~K (see eq.~\ref{dotm}). The evolution starts at the bottom left corner of the figure. A sharp increase of the  brightness temperature is caused by  HeII ionization during the formation of ionized bubble in the IGM around the growing BH. In the inner part of the bubble the  helium is ionized to HeIII, causing the $\Delta T_b$ to fall. This value reaches a maximum close to the outer radius of the HeII ionized sphere (Figure~\ref{fig-evol}). The ionization zone continues to expand and the angular size of HeII zone increases. When the accretion is  completed (this moment is marked by the  red point in Figure~\ref{fig-tb-size}), the helium ionization stops, and the luminous sphere in the $^3$HeII line is stalled at its maximum size. At this time, the  doubly ionized helium starts to recombine. Due to the  decrease of gas temperature and growth of HeII fraction the brightness temperature begins to  increase, which leads an afterglow emission of the IGM in the $^3$HeII hyperfine line. But as  the gas  cools further, the brightness  temperature falls again because 
HeII begins to recombine. 

The duration of this phase is determined by HeIII/HeII recombination kinetics and appears to be  several tens of Myrs. It is close to the accretion time for values of  $n_0$ and $T_0$ adopted here. In the case of more rapid accretion rate the luminous phase becomes longer than accretion time. Moreover, the increase of accretion rate, for instance, due to the increase of $n_0$ from $10^4$ to $2\times 10^4$~cm$^{-3}$, leads to higher ${\rm max}[\Delta T_b]$: it rises from 0.2 to 0.25~$\mu$K. In this case the accretion episode completes earlier, $z\simeq 18.6$, while the  $^3$HeII afteglow lasts longer than the accretion time. The angular size  of the luminous sphere is the same. On the contrary, a  decrease in the   accretion rate, e.g. for $n_0=5\times 10^3$~cm$^{-3}$, produces a lower signal of around 0.15~$\mu$K. 

The BHs that start growing at higher redshifts produce  a stronger signal because of higher background density. For example, the maximum signal for a BH that  started its growth at $z_0=30$ is about a factor 2 higher than that for a BH starting at $z_0 = 20$ (see Figure~\ref{fig-tb-size}, right panel). One can see that this maximum is reached at a higher redshift: $z\sim 22.5$ for  $z_0 = 30$ as compared to  $z\sim 16$ for  $z_0 = 20$. 

After the accretion on to a BH turns off helium begins to recombine not only in the intergalactic medium, but also inside the host halo, where the gas can be highly ionized. We note that the $X_{\rm HeII} \simeq 0.5$ and $T\simeq 10^4 \, \rm K$ within a radius of nearly 300~pc during this phase and this region contains a substantial fraction of the halo's mass. As the gas density in the halo is nearly  two orders of magnitude higher than the background density (the 'top-hat' model), collisional coupling is efficient inside the halo ($y_c \gg 1$ in Eq.~(\ref{tspin})) which means $T_s \simeq T$  or  $T_s^{\rm ^3HeII} \gg T_{\rm CMB}$, which is unlike the situation in the IGM (the lowest panel of Figure~\ref{fig-evol}); this  results in an  enhancement of  the signal  (Eq.~(\ref{tbhe})). For a halo of total mass $10^{8} \, \rm M_\odot$ with the background baryon to dark matter ratio, the peak flux is $0.02 \, \rm nJy$. This is smaller than the flux from the IGM and therefore can be neglected. 

Thus, a BH that  formed and started growing at high redshift $z\sim 30$ can give rise to a source of $^3$HeII hyperfine line  emission with a brightness temperature around $T_b\sim 1\mu$K and a size of roughly an arc-minute (Figure~\ref{fig-tb-size}); in addition the line width, given by expansion across the observable region, is $\delta \nu \simeq 1 \, \rm MHz$.  The flux from such a source is $F_{\nu_0} = k_{\rm B} T_b/\lambda_0^2 \Omega$, here $\Omega \simeq \theta^2$ is the solid angle subtended by the source and $\lambda_0 = c/\nu_0$. Assuming, $z \simeq 12$ (we consider the model for $z_0 = 15$), we get $\nu_0 \simeq 700  \, \rm MHz$ and the flux $F_{\nu_0} \simeq 0.3 \, \rm nJy$. Even though $\nu_0$ lies  within the range of SKA1-MID, the flux is too small to be detectable by the deep SKA1-MID surveys in the line mode. 

It is worth emphasizing the robustness of the resulting  brightness temperature to different scenarios of  BH growth.  Fig. \ref{fig-tb-size} (right panel) shows that the brightness temperature $T_b\propto (1+z)^{2.5}$  with only weakly dependence on the parameters of  BH growth. This scaling stems from the thermodynamics of the gas in the ionized bubble. After ionization by the growing central BH gas, the  temperature increases up to $T\simeq 3\times 10^5$~K nearly independent of the model of BH accretion, and when the accretion exhausts the gas begins to cool radiatively with the rate $\Lambda(T)\propto T^2$. With gas density $n\propto (1+z)^3$ temperature falls  as     
\be 
{dT\over dz}\propto -\Omega_m^{-3/2}T^2(1+z)^{1/2},
\ee
where $\Omega_m$ is the matter density parameter. One can further  infer approximately $T(z)\propto\Omega_m^{3/2}(1+z)^{-3/2}\Delta z$, with $z$ being the redshift when the gas started to cool. When this dependence is substituted into $y_cT$ in Eq. (\ref{tspin}), we get  $y_cT\propto (1+z)^{-2.75}$. This redshift dependence mostly  determines  the brightness temperature $T_b$, which explains the  $T_b$ vs $z$ scaling  seen in  Fig. \ref{fig-tb-size}.

We next consider whether a radio interferometer such as SKA1-MID can detect the flux owing to a collection of such $^3$HeII hyperfine line emitting  regions. We consider the parameters of SKA1-MID for our analysis. In continuum observation, we assume  a bandwidth $\Delta\nu_0 = 100 \, \rm MHz$ and a range of synthesized beams, $10\hbox{--}100$~arc-seconds,  for which the flux sensitivity of SKA1-MID is the highest and nearly the same in the frequency range of interest. The volume bounded by the solid angle of the systhesized beam and the bandwidth is $dV \simeq \Delta\nu_0 r^2dr d\Omega_{\rm syn}/d\nu_0$. The number density of haloes in the mass range $10^7\hbox{--}10^8 \, \rm M_\odot$ contribute principally to the signal. The (comoving)  number density of such haloes can be computed from Press-Schcheter formalism:  $Mdn/dM \simeq 50 \, \rm Mpc^{-3}$ at $z\simeq 12$. The total number of $^3$HeII hyperfine line emitting regions in the volume are  $N\simeq \eta_{\rm bh} Mdn/dM dV$, where $\eta_{\rm bh}$ is the fraction of haloes that host BHs. The total flux from all the sources in the volume $N F_{\nu_0}$ and its fluctuation, assuming randomly spatial distribution of  sources, is $N^{1/2} \eta_{\rm bh}  F_{\nu_0}$. Being an interferometer, SKA1-MID can only detect the fluctuation  of the signal. We get:
\ba
S_{\rm ^3HeII}/\eta_{\rm bh}  \simeq N^{1/2} F_{\nu_0} \simeq 150\, {\rm nJy} \left ({T_b \over 1 \, \rm \mu{\rm K}} \right ) \times \\ \nonumber
 \times \left (\theta \over 0.5' \right)^2 \left (\Omega_{\rm syn} \over 1.5' \right )^{1/2} \left (\Delta\nu \over 100 \, \rm MHz \right )^{1/2}
\ea

For the parameters used above, SKA1-MID deep continuum survey can reach an RMS of nearly 150~nJy in 1000~hours of integration\footnote{https://www.skatelescope.org/wp-content/uploads/2014/03/SKA-TEL\_SCI-SKO-SRQ-001-1\_Level\_0\_Requirements-1.pdf}, which suggests that the $^3$HeII signal is in the range  of  SKA1-MID but it would be a challenge to detect it.

In the foregoing, we considered a single accretion episode. It is possible the BH could undergo many such episodes. We note that multiple episodes of BH growth with accretion periods shorter than HeIII recombination time would diminish the brightness temperature in  the hyperfine  line, while increasing the angular size of the emitting region. In such a case, the evolutionary tracks on the angular size--brightness temperature plane (Figure~\ref{fig-tb-size}) would consist of multiple loops of increasing $\theta$ but diminishing brightness temperature.

{
In the foregoing we  assumed that the only source of ionization and heating of the circumgalactic and intergalactic medium is radiation produced by a hot super-critical accretion disk of the growing central BH. In this regard the other energy source -- stellar activity in the host galaxy is worth mentioning. The most powerful  energy reservoir associated with stellar population, particularly at the early epochs $z\sim 10\hbox{--}20$, are high-mass X-ray binaries. For large X-ray luminosity $L_X>3\times 10^{40}$ erg~s$^{-1}$,  $L_X$ and star formation rate follow a  relation  \citep{grimm03,grimm04}} :
\be 
L_X\simeq 6.7\times 10^{39}{{\rm SFR}\over \msun~{\rm yr}^{-1}}~{\rm erg~s^{-1}}.
\ee
{
In the cases we study, an accreting  BH with the initial mass $M_{BH}(z_0=20)=300\msun$ grows by a factor of 1000 by  $z=17.2$,  which causes its bolometric luminosity to  increase from $\sim 3\times 10^{41}$ to $\sim 3\times 10^{44}$ erg~s$^{-1}$, as follows from Eq. (\ref{edreg}). It is seen that the  star formation may compete with the accreting BH in the generation of X-ray photons only for  an exceptionally high SFR$\sim 10^3\hbox{--}10^6\msun~$yr$^{-1}$. It allows us to conclude that the  contribution of stellar sources to the X-ray radiation in the IGM  surrounding a grow BH is expected to be negligible \citep[for details  on the contribution of stellar sources see ][]{vrss18}.
}

{
We could also address the following question: what is the impact of distant sources on the ionization state of Helium surrounding an accreting BH. It can readily be shown that photons of energy less than a few hundred eV are locally absorbed which we have taken into account in our modelling. Photons for much larger energy (e.g. soft x-ray) can penetrate regions far away from their source of origin. It has been shown by \citet{steenberg85}
that these photons make a negligible contribution to the ionized state of Helium especially if the ionization level is already high, as is our case where the hydrogen  is almost fully ionized. Most of this energy goes into heating the medium to a temperature which is generally smaller than $10^4$, the temperature of the gas in our case. This means such photons will have negligible impact of either ionization or the thermal state of the gas. This is in agreement with   the  detection of singly ionized helium at $z\simeq 3$ in the IGM, which shows that the second ionization of the helium was delayed with respect to the ionization of hydrogen and the first ionization of the helium \citep[e.g.][]{2019ApJ...875..111W,1994Natur.370...35J}.
}

{
In our study we assume that the CMB is the only radio source at high redshifts. This allows us to reach the conclusion that the  hyperfine line from $^3$HeII is detectable in emission when the spin temperature exceeds the CMB temperature. This is also an important difference between our case and the local detection of the hyperfine line \citep[e.g.][]{2018AJ....156..280B,2010IAUS..268...81B,2006ApJ...640..360B}. We note that if there are additional radio sources much brighter than the CMB at redshifts of interest to us in this paper,  the signal could also be seen in absorption.
}

\section{Conclusions}  \label{sec:sumcon}

In this paper, we explore the possibility of measuring the BBN $^3$He abundance by the detection of $^3$HeII hyperfine line 8.67~GHz emission from   ionized zones around accreting  black holes that form at high redshifts, $z=15\hbox{--}30$. We model the ionization and thermal evolution of both the halo surrounding the BH and the expanding IGM surrounding the halo. 

We compute the brightness temperature and the angular extent of regions emitting in the $^3$HeII hyperfine in the IGM. We also estimate the flux in the line from the halo. The $^3$HeII hyperfine line emission from the IGM reaches a maximum  after the accretion onto the BH is  finished. After this phase, the HeIII zones around the BH begin to recombine and produce higher HeII fraction, which reaches up to $\sim 0.5$ of the total He abundunce. This luminous 'afterglow' phase lasts over 40~million years if the accretion ends at $z \simeq 17$.  Its duration is  longer  if the redshift at which the accretion ends is lower.

We find that the maximum brightness temperature is $\simeq 0.2 \mu$K to $\simeq$0.5~$\mu$K, depending on the epoch when such a black hole starts growing, and the maximum angular size is around $0.5'$. The flux from such a region is $\simeq 0.3 \, \rm  nJy$. The flux from the { halo hosting the BH} is an order of magnitude smaller. We investigate the detection of this signal using SKA1-MID. While the flux from a single region is too small to be detected, we estimate the expected flux from collection of such regions  for a volume bounded by the  synthesized beam and the bandwidth of SKA1-MID. The RMS of this quantity is detectable by SKA1-MID. We find that while it would still be a challenge to detect this signal, it is within the range of expected 
sensitivity of SKA1-MID.  

\section{Acknowledgements}

\noindent

This work is supported by the joint RFBR-DST project (RFBR 17-52-45063, DST P-276). EV is grateful to the Ministry for Education and Science of the Russian Federation (grant 3.858.2017/4.6). The work by YS is done under partial support from the joint RFBR-DST project (17-52-45053), by the project 01-2018 ``New Scientific Groups LPI'', and the Program of the Presidium of RAS (project code 28).


\end{document}